# Bhirkuti's Relative Efficiency (BRE): Examining its Performance in Psychometric Simulations

Aneel Bhusal[1], Todd Little[1,2]

[1]Department of Educational Psychology, Leadership, and Counseling, Texas Tech University

[2]Optentia Research Focus Area, North-West University

Correspondence concerning this article should be addressed to Aneel Bhusal, Email: abhusal@ttu.edu

## Abstract

Traditional Relative Efficiency (RE), based solely on variance, has limitations in evaluating estimator performance, particularly in planned missing data designs. We introduce Bhirkuti's Relative Efficiency (BRE), a novel metric that integrates precision and accuracy to provide a more robust assessment of efficiency. To compute BRE, we use interquartile range (IQR) overlap to measure precision and apply a bias adjustment factor based on the absolute median relative bias (AMRB). Monte Carlo simulations using a Latent Growth Model (LGM) with planned missing data illustrate that BRE maintains theoretically consistency and interpretability, avoiding paradoxes such as RE exceeding 100%. Visualizations via boxplots and ridgeline plots confirm that BRE provides a stable and meaningful estimator efficiency evaluation, making it a valuable advancement in psychometric and statistical modeling. By addressing fundamental weaknesses in traditional RE, BRE provides a superior, theoretically justified alternative for relative efficiency in psychometric modeling, structural equation modeling, and missing data research. This advancement enhances data-driven decision-making and offers a methodologically rigorous tool for researchers analyzing incomplete datasets.

# Introduction

## Overview of Relative Efficiency (RE) in Missing Data Research

In statistical and psychometric research, Relative Efficiency (RE) has long been a standard metric for comparing estimator performance under missing data conditions. Traditional RE is computed as the ratio of variances between estimates obtained from a reference group (complete dataset) and those from a comparison group (incomplete dataset) processed using a missing data method. A higher RE value suggests minimal efficiency loss due to missing data, implying that the estimation approach is retaining most of the information available in the complete data. RE has been widely used in planned missing data designs, where missingness is introduced systematically to optimize data collection and reduce participant burden (Graham et al., 2001; Rhemtulla et al., 2014). The approach is particularly relevant in Full Information Maximum Likelihood (FIML) estimation and Multiple Imputation (MI) methods, where the goal is to derive efficient and unbiased parameter estimates despite the presence of missing values. Despite its broad application, however, variance-based RE suffers from critical limitations, which often lead to misinterpretations of estimator performance.

## Limitations of RE: Variance Inversion, Misleading Interpretations, and Neglect of Bias

While RE is a useful measure of efficiency in ideal conditions, it is subject to several methodological flaws that undermine its reliability as a stand-alone metric. The three primary concerns with traditional RE are:

***Variance Inversion and Instability.*** One of the most problematic aspects of RE is its sensitivity to variance inversion. Ideally, the variance of an estimator derived from a reference group should be lower than that of a comparison group. However, in certain cases, particularly in small sample conditions or when outliers are present, the variance of an incomplete dataset comparison group may, by chance, be lower than that of the complete dataset reference group. This leads to RE values exceeding 100%, creating a statistical paradox where an estimator based on incomplete data appears more efficient than one based on the complete dataset. As an example, during iterative simulation processes, variance inversion can occur due to the random outliers. If the reference group contains a higher frequency of outliers, its variance becomes inflated, potentially leading to a misleading perception of greater precision in the incomplete dataset. This results in an overestimated RE (greater than 100%), suggesting that the comparison estimator is more efficient. Conversely, if outliers are more prevalent in the incomplete dataset, its variance is artificially inflated, leading to a lower RE and an underestimation of the estimator's true efficiency. Such a scenario implies that the incomplete dataset outperforms the complete dataset in terms of variance, contradicting theoretical expectations and leading to misleading efficiency interpretations. Such anomalies arise due to uneven sample distributions, differences in admissible solutions, or extreme data values that distort variance-based comparisons (Wu et al., 2016).

***Misleading Interpretations***. Because traditional RE is based purely on variance, it fails to consider accuracy. An estimator may exhibit lower variance but still be biased, meaning its expected value deviates systematically from the true population parameter. In psychometric modeling, an estimator with low variance but high bias may yield misleading conclusions. For instance, two estimation methods might have identical variance, yet one could systematically overestimate or underestimate key parameters. Without incorporating bias correction, RE may provide an overly optimistic assessment of an estimator's performance.

***Failure to Account for Bias in Efficiency Calculations.*** Traditional RE calculations assume that variance alone determines efficiency, neglecting bias as a crucial component of estimation accuracy. Bias is particularly problematic in missing data contexts where non-random patterns of missingness (MNAR) may introduce systematic deviations. Without an adjustment for bias, RE can overstate the efficiency of methods that introduce estimation errors, leading to incorrect methodological decisions.

**Introduction of Bhirkuti's Relative Efficiency (BRE) as a Robust Alternative**

To address the shortcomings of traditional RE, this study introduces Bhirkuti's Relative Efficiency (BRE) a novel efficiency metric that integrates both precision and accuracy. Unlike RE, which relies solely on variance comparisons, BRE is formulated using both precision and accuracy. Precision (Variance Component) is quantified via Interquartile Range (IQR) overlap, ensuring that the estimator's stability and consistency are reflected. Accuracy (Bias Component) is adjusted using the Absolute Median Relative Bias (AMRB) to correct for systematic estimation errors.

This dual-component formulation ensures that efficiency estimates remain theoretically valid, preventing paradoxical cases where missing data estimators appear more efficient than the complete dataset. By incorporating both distributional similarity (via IQR overlap) and bias correction, BRE provides a more meaningful and interpretable assessment of estimator performance. The objective of this study is to demonstrate that Bhirkuti's Relative Efficiency (BRE) provides a more robust and interpretable measure of relative efficiency than variance-based RE. Through a series of Monte Carlo simulations, we will compare traditional RE and BRE across multiple planned missing data conditions, including different levels of missingness, differing magnitude, sample sizes, and estimation methods. We will evaluate the stability of BRE using visual analyses (boxplots and ridgeline plots) to confirm its consistency. Finally, we will validate BRE as a superior alternative by showing that it remains theoretically sound, particularly in scenarios where traditional RE fails. By addressing the inherent weaknesses of variance-based relative efficiency measures, this study establishes Bhirkuti's Relative Efficiency as a significant methodological advancement in psychometric modeling, statistical estimation, and missing data research.

## Traditional Relative Efficiency (RE)

Relative Efficiency (RE) is a widely used metric for evaluating the performance of an estimator in the presence of missing data. It is traditionally defined as the ratio of the variance of parameter estimates from a reference group to the variance of estimates from a comparison group, given by:

$$RE_{\hat{\theta}} = \frac{var(\hat{\theta})_{reference\ group}}{var(\hat{\theta})_{comparision\ group}} * 100\% \quad \ldots\ldots\ldots\ldots \textit{(1)}$$

Where $\hat{\theta}$ is a given parameter estimate of interest.

This formulation assumes that a complete dataset provides the most precise estimates, and any missingness introduced should theoretically reduce efficiency. RE values closer to 1.0 indicate minimal efficiency loss, suggesting that the missing data-handling technique retains a high proportion of the information available in the complete data. An RE of 0.80, for example, implies that the estimates obtained from the incomplete data are as efficient as those derived from a dataset with 80% of the original sample size (Muthén et al., 1987; Garnier-Villarreal et al., 2014; Rhemtulla et al., 2014; Wu et al., 2016).

Despite its intuitive appeal, however, RE has critical limitations that challenge its reliability as a performance metric. One major issue is that RE values can exceed 100%, leading to paradoxical interpretations where incomplete data appear to yield more precise estimates than complete data. This paradox typically occurs in cases of variance inversion, where the variance of estimates from the incomplete dataset is unexpectedly lower than that of the complete dataset. Such results often stem from factors such as unequal sample distributions, extreme values, or inadmissible solutions, rather than genuine improvements in efficiency.

## Bhirkuti's Relative Efficiency (BRE): A Robust Alternative

To address the limitations of traditional Relative Efficiency (RE), this study introduces Bhirkuti's Relative Efficiency (BRE) a novel metric designed to integrate both precision and accuracy in estimator performance evaluation. Unlike RE, which is solely based on variance comparisons, BRE incorporates two critical components, precision and accuracy. The precision component is measured using the Interquartile Range (IQR) Overlap, which assesses the degree of similarity between the distributions of estimates from the reference and comparison group. This overlap provides a more robust and stable measure of efficiency, unaffected by extreme values and variance distortions. The accuracy component is achieved by using the Absolute of Median Relative Bias (AMRB) of the comparison group, which corrects systematic deviations between estimated and true parameter values. By accounting for bias, BRE ensures that efficiency calculations reflect not only the precision similarity but also their closeness to the true population values. The interquartile range (IQR) is derived from the entire dataset, ensuring that no data points are excluded in its calculation. By capturing the spread of the central 50% of observations, it provides a robust measure of dispersion that accurately reflects the true characteristics of the distribution while minimizing the influence of extreme values. BRE is formulated as follows:

*Bhirkuti's Relative Efficiency (BRE)*

$BRE = IQR\ Overlap * (1 - |Median\ Relative\ Bias|)$.…………………*(2)*

**Component 1: Precision in Estimation.**

***General Case: Precision Measurement Using Interquartile Range (IQR) Overlap***

IQR Overlap, inspired by Dice Similarity (Dice, 1945), quantifies estimator precision by measuring the relative intersection of their interquartile ranges, ensuring interpretability and robustness. This robust metric assesses estimator precision through the proportional overlap of their interquartile ranges, providing a consistent and interpretable measure across different distribution.

$IQR\ Overlap_{Comparision,Reference} = \frac{2*Intersection_{Comparision,\ Reference}}{Sum\ of\ IQRS_{Comparision,\ Reference}}$ …………….*(3)*

Where,

$$\text{Intersection}_{\text{Comparision,Reference}} = \min(Q3_{Comparision}, Q3_{Reference}) - \max(Q1_{Comparision}, Q1_{Reference})$$

$$Sum\ of\ IQRS\ _{FIML,Complete\ Dataset} = (Q3_{Comparision} - Q1_{Comparision}) + (Q3_{Reference} - Q1_{Reference})$$

***Special Case: Precision Measurement When the Comparison Estimator's IQR is Nested Within the Reference IQR***

When the comparison estimator's IQR is narrower and entirely nested within the reference estimator's IQR, meaningful precision assessment in this scenario**:**

$IQR\ Overlap_{Reference,\ Comparision} = \frac{IQR_{Reference}}{IQR_{Comparision}}$ ……………………*(4)*

**Component 2: Bias in estimation.**

***Median Relative Bias: Systematic Estimation Deviations***

Median Relative Bias represents the systematic bias in estimation, ensuring that efficiency values are adjusted for accuracy.

$$Median\ Relative\ Bias = \ Median\ \left(\frac{\hat{\theta}_1-\theta}{\theta}, \frac{\hat{\theta}_2-\theta}{\theta}, \ldots.., \frac{\hat{\theta}_n-\theta}{\theta}\right) \ldots\ldots\ldots\ldots\ldots\ldots(5)$$

Where:

$\theta$= true population parameter value;

$\hat{\theta}_n$ = estimated parameter value across all converged replications for a given condition (Collins, Shafer, & Kam, 2001; Graham, 2009).

This dual-component structure gives BRE several theoretical advantages over RE. First, it accounts for both variance and bias, providing a more comprehensive measure of estimator performance. Second, it eliminates misleading efficiency estimates, ensuring that missing data techniques do not appear paradoxically more efficient than complete data and remain stable across different missing data scenarios, including cases with variance inversion, small sample sizes, and outliers.

While the selected reference group may contain some level of bias, the primary objective is to assess the efficiency of the comparison group relative to the reference group. If evaluating the efficiency of the reference group itself, its own bias would be incorporated into the adjustment process. Because Bhirkuti's Relative Efficiency (BRE) is designed to evaluate a specific estimator (comparison group), the bias correction is applied based on its median relative bias. This approach ensures that BRE provides an efficiency estimate that accurately reflects the performance of the chosen estimator while maintaining theoretical consistency. By treating the selected reference dataset as the gold standard, BRE remains aligned with established efficiency frameworks, allowing for meaningful comparisons across different estimation conditions. By integrating distributional similarity (through IQR Overlap) with bias correction, BRE overcomes the fundamental flaws of traditional RE, making it a more reliable, interpretable, and theoretically sound efficiency metric for evaluating missing data estimation techniques. A key feature of Bhirkuti's Relative Efficiency (BRE) is its ability to remain interpretable and robust across different estimator distribution.

## Simulation Study

To evaluate the performance of Bhirkuti's Relative Efficiency (BRE) in comparison to traditional Relative Efficiency (RE), this study employs a Monte Carlo simulation (Carsey & Harden, 2013) using a Latent Growth Curve Model (LGM) (McArdle et al., 1987). The simulation is designed to examine how planned missing data affects parameter estimation and to determine whether BRE provides a more stable and interpretable efficiency metric.

### Latent Growth Curve Model (LGM)

The simulated data is based on a bivariate Latent Growth Model (LGM) with two psychological constructs (Hertzog et al., 2006). Bullying (B), a latent construct measuring the frequency of bullying behaviors, and Homophobic Teasing (H), a latent construct assessing the prevalence of homophobic teasing incidents. Both constructs are measured at five equally spaced time points, allowing for the assessment of developmental changes over time. Each construct is modeled using three observed indicators per time point, ensuring adequate measurement precision while maintaining model parsimony. The data generation process follows the structure outlined in Rhemtulla et al. (2014) and Little (2024), ensuring consistency with prior methodological research on missing data estimation. The underlying population parameters, including growth trajectory means, variances, and covariances, are derived from empirical studies in longitudinal developmental psychology by Rhemtulla et al. (2014).

Figure 1

Proposed Growth Curve Model for Simulation

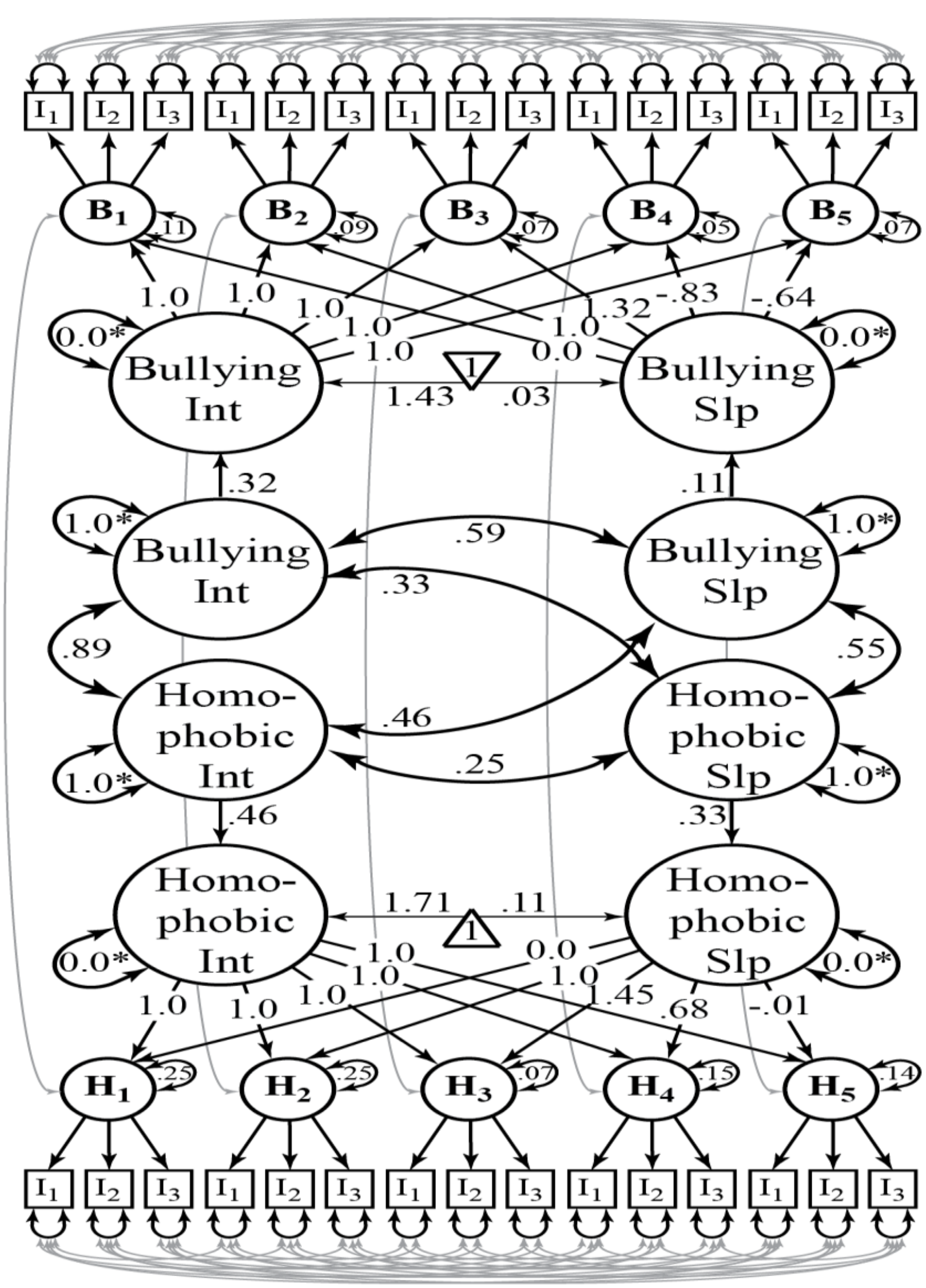

Note: Figure 11.7, in Longitudinal Structural Equation Modeling (p. 398), by T. D. Little, 2024, Guilford Publications. Copyright 2024 by T. D. Little. Reprinted with permission.

The simulation systematically introduces planned missing data patterns, testing whether BRE accurately captures both precision and bias-adjusted efficiency under different conditions. By implementing this structured LGM framework, the study ensures that findings are both theoretically grounded and empirically robust, reinforcing the validity of BRE as an alternative efficiency metric.

**Simple Wave Missing Design (SWMD-6)**

To systematically examine the impact of planned missing data on estimator performance, this study employs the Simple Wave Missing Design (SWMD-6), a widely used planned missingness structure originally proposed by Graham et al. (2001). SWMD-6 is designed to optimize data collection while maintaining statistical power and estimator efficiency in longitudinal research. The SWMD-6 consists of six distinct participant groups, each following a unique missingness pattern across five measurement occasions. While one group (Group 1) is observed at all time points (complete data condition), the remaining five groups (Groups 2–6) have systematically missing observations at specific time points, following an efficiently structured missing data pattern.

Table 1

Simple Wave Missing Design with Six Groups (SWMD-6)

| Random | Occasion Of Measurement | | | | |
|---|---|---|---|---|---|
| Group | Time 1 | Time 2 | Time 3 | Time 4 | Time 5 |
| 1 | √ | √ | √ | √ | √ |
| 2 | √ | √ | √ | √ | X |
| 3 | √ | √ | √ | X | √ |
| 4 | √ | √ | X | √ | √ |
| 5 | √ | X | √ | √ | √ |
| 6 | X | √ | √ | √ | √ |

Note: Within each time occasion, √ = data present, and X = data missing; (Graham et al., 2001; Little, 2024)

This missing data structure ensures that each time point remains well-represented across participants, preventing severe information loss while introducing planned missingness for efficiency. The staggered missing data pattern allows for robust statistical inference while reducing participant burden, making it a widely adopted design in psychometric and developmental research. One of the key advantages of SWMD-6 is its ability to maintain model convergence in Full-Information Maximum Likelihood (FIML) estimation. FIML leverages all available data points to estimate model parameters, making it particularly well-suited for planned missing designs (Rhemtulla et al., 2014). By implementing SWMD-6, this study ensures that Bhirkuti's Relative Efficiency (BRE) can be systematically evaluated across multiple planned missingness conditions, reinforcing the metric's robustness and theoretical validity in longitudinal research settings.

**Full-Information Maximum Likelihood (FIML) Estimation**

Full-Information Maximum Likelihood (FIML) is a widely used statistical approach for handling missing data by directly estimating model parameters using all available data points without the need for explicit imputation. Unlike multiple imputation (MI), which generates complete datasets by replacing missing values with plausible estimates, FIML computes likelihood functions for each observed case, maximizing the probability of the given data structure. This approach allows for more efficient parameter estimation while maintaining statistical rigor, making it particularly useful in structural equation modeling and longitudinal analyses (Dempster et al., 1977; Enders et al., 2001; Enders, 2022). FIML performs optimally when data are missing completely at random (MCAR) or missing at random (MAR), as it provides unbiased parameter estimates under these conditions. The MAR assumption ensures that the probability of missingness can be fully explained by observed variables, allowing FIML to recover valid estimates using the available data (Enders, 2022).

**Simulation Design and Conditions**

To rigorously assess the performance of Bhirkuti's Relative Efficiency (BRE) in comparison to traditional variance-based Relative Efficiency (RE), a series of Monte Carlo simulations were conducted under systematically varied conditions. The simulations manipulated latent slope correlations, sample sizes, SWMD-6 and complete dataset patterns to evaluate estimator performance across diverse analytical scenarios. Specifically, latent slope correlations ($\rho_{s1,s2}$) were set at three levels: 0.1, 0.3, and 0.55, capturing a range of weak to moderate relationships between latent growth trajectories. Sample sizes for each of the six groups varied from small (n = 40, 60, 80, 100) to moderate (n = 300, 500) and large-scale conditions (n = 800, 1000), allowing for an assessment of estimator efficiency and bias across different levels of statistical power. FIML was applied to SWMD-6 to obtain parameter estimates for evaluating estimator performance. Each condition was replicated 5,000 times, ensuring robust parameter estimation and stable performance metrics. The complete dataset served as the reference group, while the FIML estimator applied to SWMD-6 functioned as the comparison group for evaluating estimator performance.

The evaluation of estimator performance was conducted using three key statistical criteria. First, Relative Bias (RB) was computed to quantify systematic deviations between estimated and true parameter values (Muthén et al., 1987; Garnier-Villarreal et al., 2014; Lang et al., 2020; Mcardle & Epstein, 1987; Moore et al., 2020). Second, Relative Efficiency (RE) was compared against the proposed Bhirkuti's Relative Efficiency (BRE) to determine the stability and reliability of the efficiency estimates, particularly in conditions where variance-based traditional RE exhibited inconsistencies. Finally, graphical analyses, including boxplots and ridgeline plots, were employed to visualize the distribution of estimates across conditions, providing an intuitive representation of estimator precision and accuracy. These visual tools allowed for a clearer interpretation of performance trends and help validate the theoretical advantages of BRE in psychometric simulations. All simulation and analysis were conducted in the R/RStudio environment using packages such as "lavaan" and "ggplot2".

## Results

### Traditional Relative Efficiency (RE) vs. Bhirkuti's Relative Efficiency (BRE)

Relative Efficiency (RE), when computed solely based on variance, exhibits significant limitations in small sample conditions and in the presence of extreme values. As shown in Figure 2, variance-based RE fails to provide reliable efficiency estimates when sample sizes are small, often producing inflated values that exceed 100%, contradicting theoretical expectations. This phenomenon, known as variance inversion, occurs when the variance of the incomplete data is lower than that of the complete dataset due to outliers, estimation artifacts, or high variability in missing data patterns. The decreasing RE trend with increasing sample size further confirms that variance-based RE is not a theoretically sound approach for assessing efficiency. Variance-based RE leads to misleading interpretations, often making incomplete data appear more efficient due to variance distortions. This instability results in unreliable efficiency estimates, failing to accurately reflect estimator performance across different conditions.

Figure 2

Comparison between Bhirkuti's Relative Efficiency (BRE) vs Traditional RE

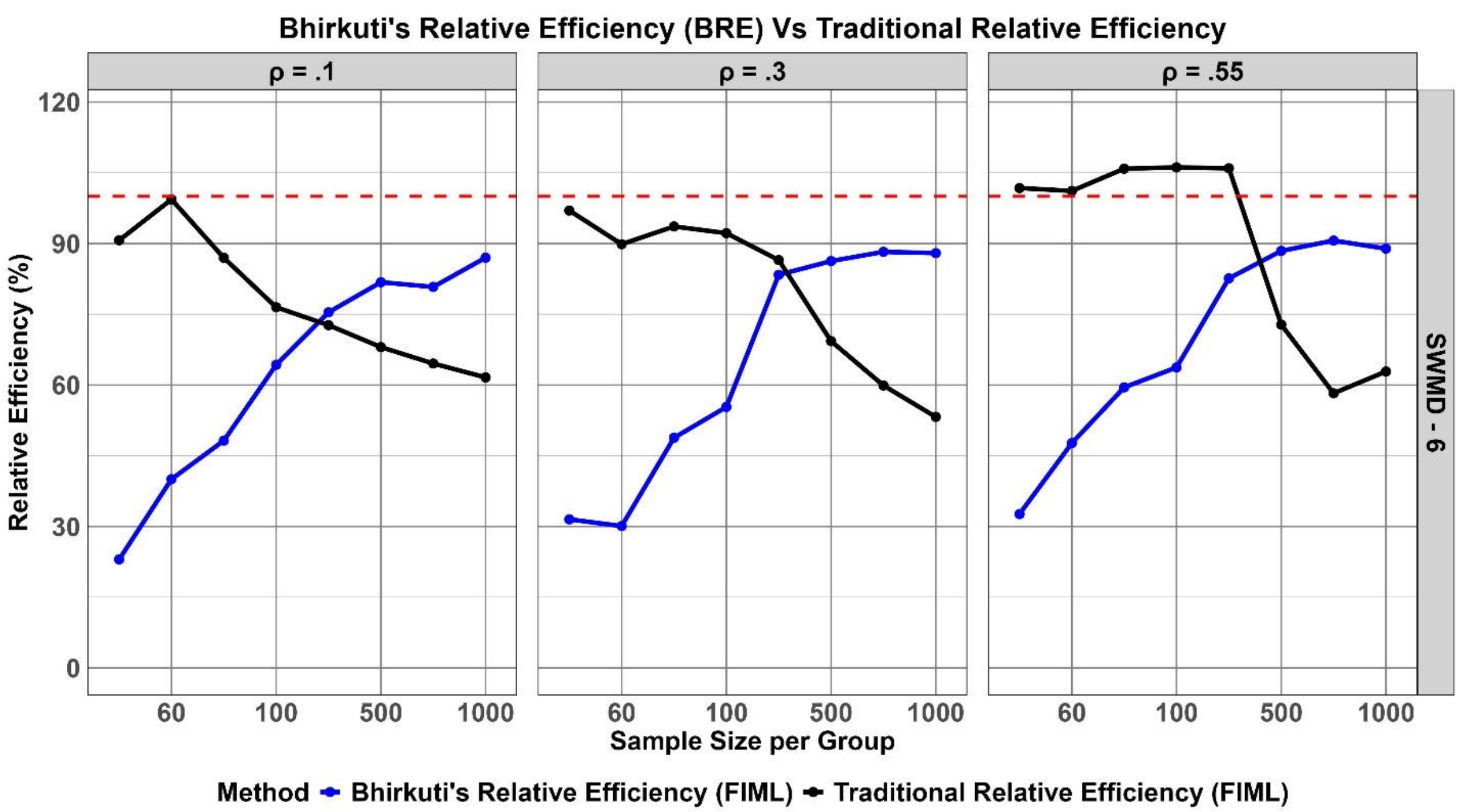

***Note:*** Red dashed horizontal line represents the 100% relative efficiency level.

As seen in Figure 2, Bhirkuti's Relative Efficiency (BRE) provides a robust and theoretically consistent efficiency metric by integrating both precision and accuracy components. As sample size increases, efficiency is expected to improve due to reduced sampling variability and more stable parameter estimates, aligning with established findings on planned missing data designs (Graham et al., 2001; Enders, 2022; Little, 2024). As demonstrated in Figure 2, BRE remains stable across all sample sizes, changing correlation and missing data conditions, mitigating the distortions caused by variance-based RE. Unlike RE, which can overstate efficiency when variance inversion occurs, BRE ensures that efficiency values remain within interpretable bounds, making it a more reliable tool for evaluating estimator performance.

**Visualization of Estimator Performance**

To further illustrate the shortcomings of variance-based RE and the advantages of BRE, we present a series of graphical analyses. Boxplots and ridgeline plots provide additional validation of these findings, demonstrating the distribution of estimates across simulated conditions.

Figure 3. Ridgeline Plot

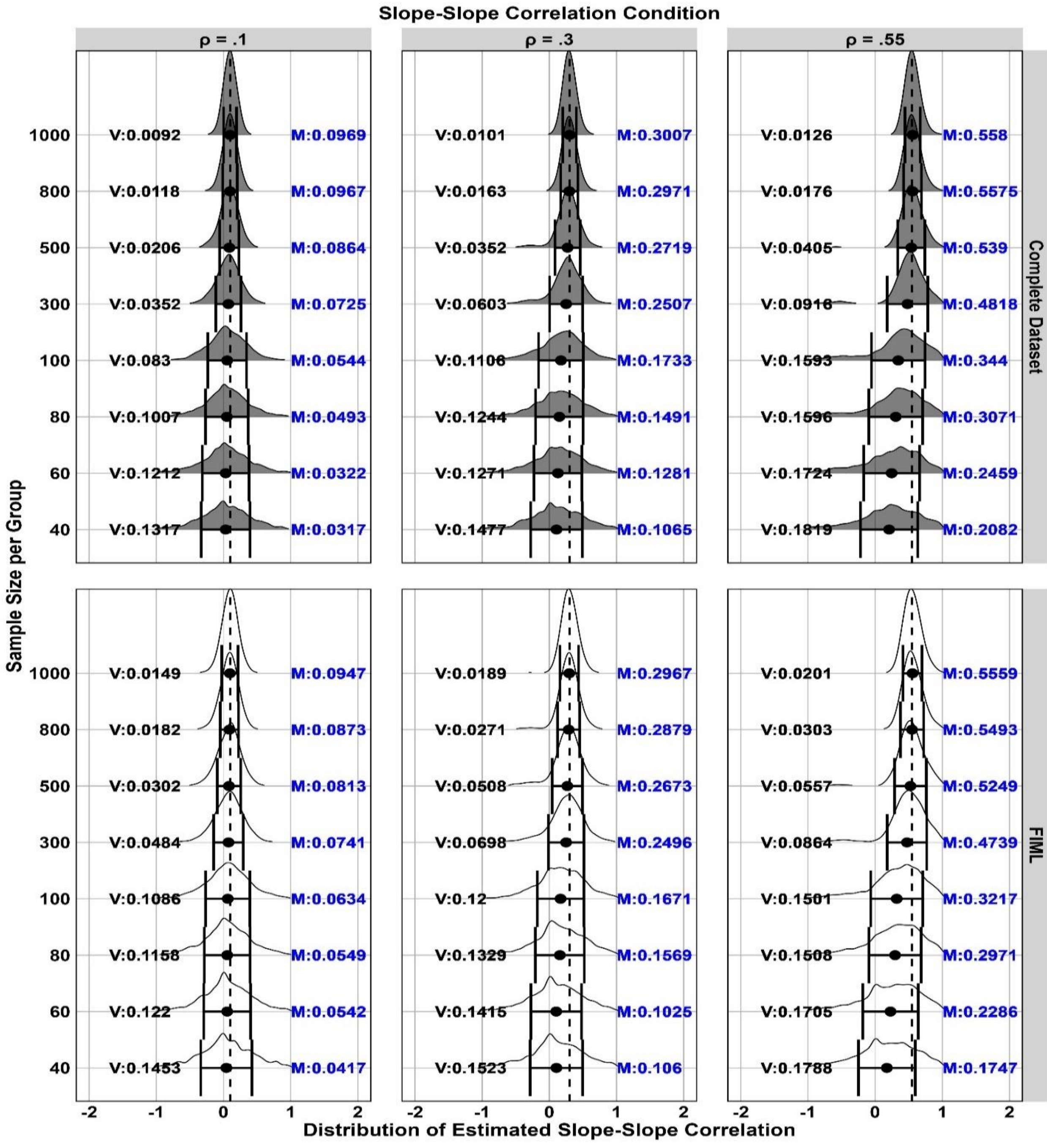

***Note:*** Vertical dashed line represents the true population value of the slope-slope correlation. The black dots indicate mean of the estimated correlations, while the error bars represent ±1 standard deviation around the mean. "V:" indicates variance. "M:" indicates mean.

Figure 4: Box-plot along with rigid-line distribution curves for Slope-Slope Correlation

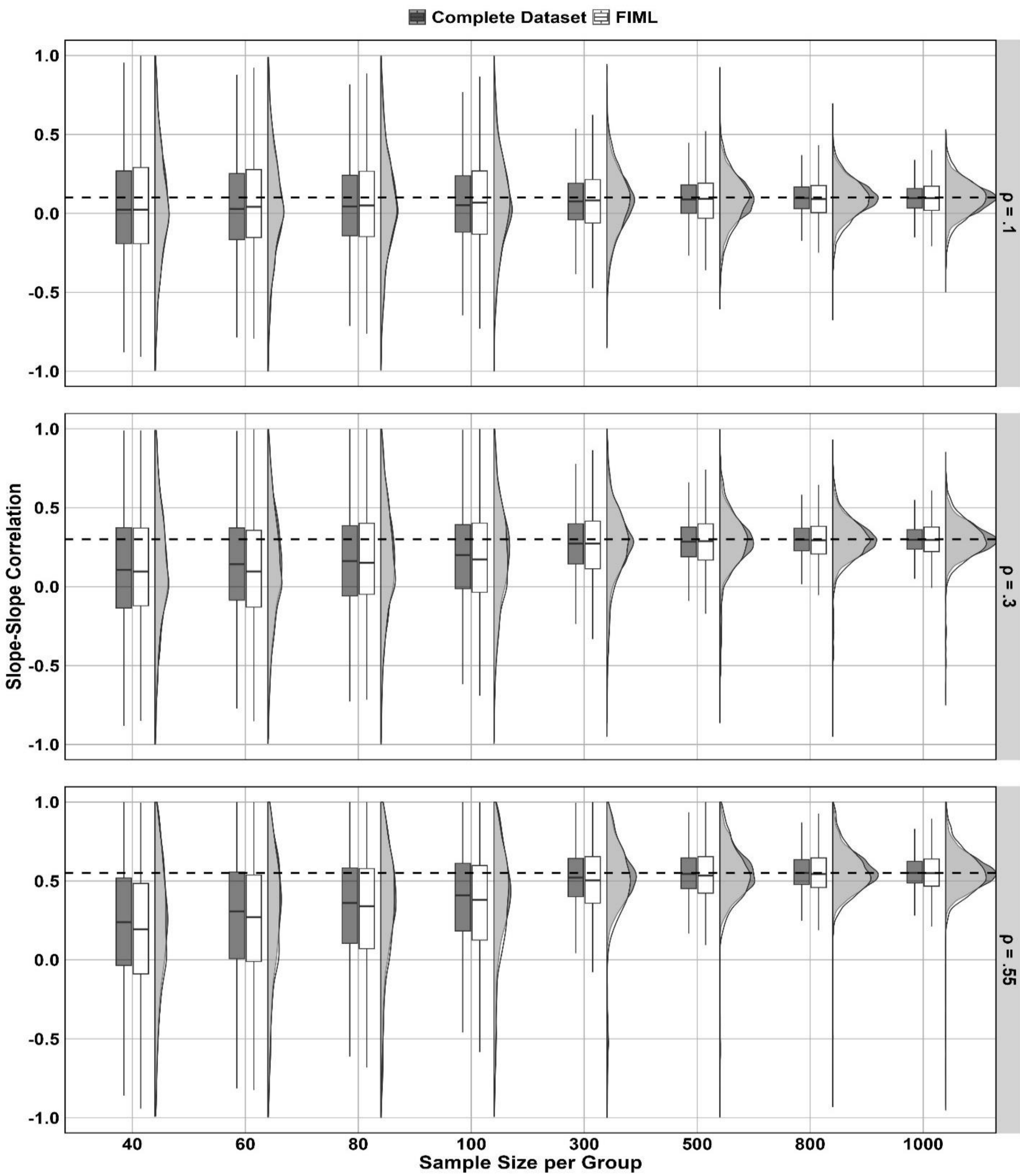

Note: Horizontal dashed line represents the true population value of the slope-slope correlation. Boxplots represent the distribution of estimated slope-slope correlations, with the central line indicating the median and the boxes capturing the interquartile range. Rigid-line density curves illustrate the shape of the estimate distributions.

Figure 5: Box-plot along with rigid-line distribution curves for Relative Bias of Slope-Slope Correlation

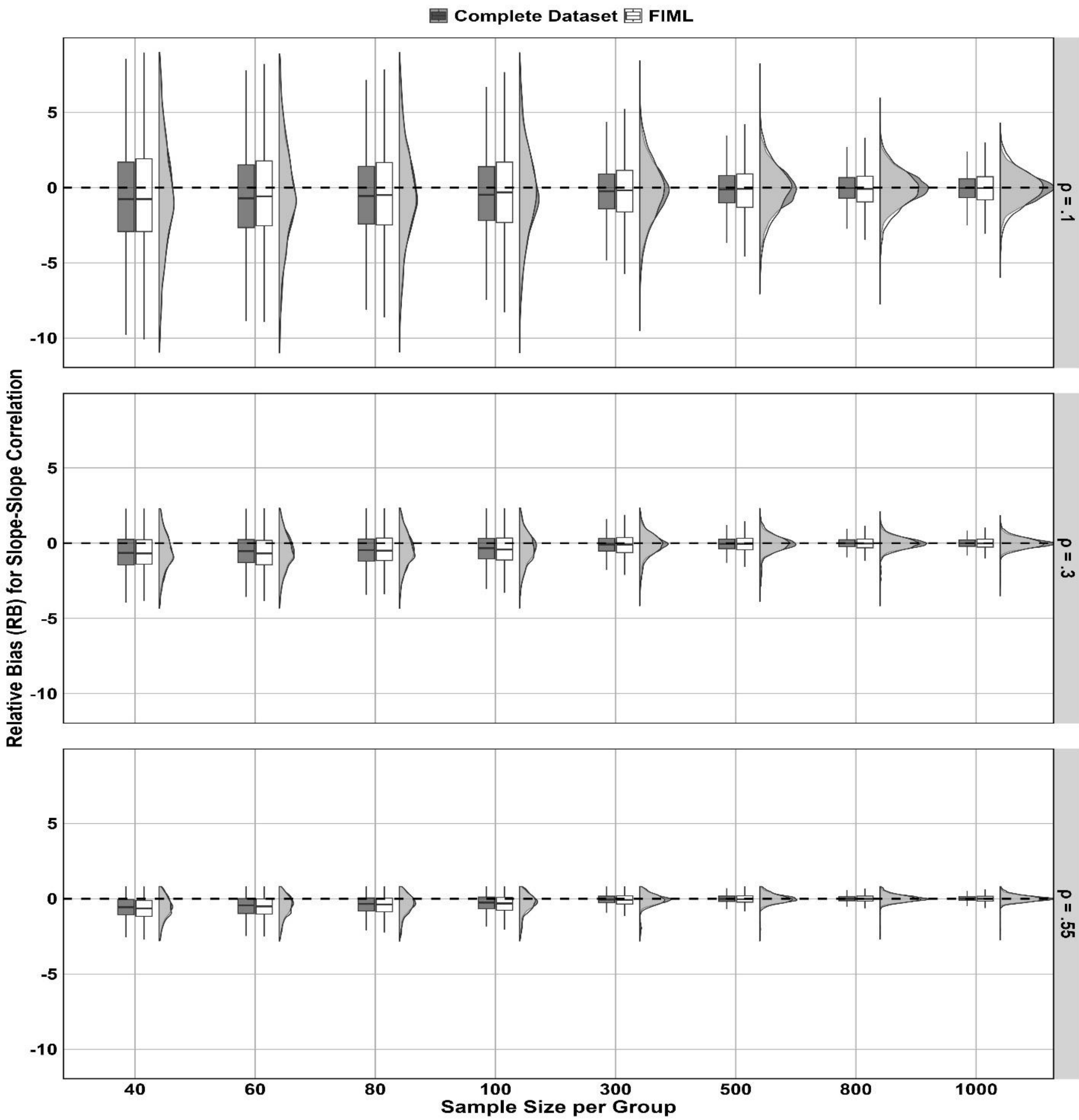

***Note:*** Horizontal dashed line at zero represents the aspiration for unbiased estimates. Boxplots represent the distribution of relative bias (RB) estimates for slope-slope correlation, with the central line indicating the median and the boxes capturing the interquartile range. The rigid-line density curves illustrate the shape of the bias distribution.

Figures 3, 4, and 5 illustrate the limitations of variance-based Relative Efficiency (RE) and the advantages of Bhirkuti's Relative Efficiency (BRE) in evaluating estimator performance under missing data conditions. Figure 3, a ridgeline plot, visually captures the distribution of slope-slope correlation estimates for both FIML and the Complete Dataset. The spread of these distributions highlights how RE exhibits greater variability and instability, particularly in conditions where variance inversion occurs when the incomplete dataset has lower variance than the complete dataset, leading to inflated or misleading efficiency values. This issue arises because RE is purely variance-based and does not consider whether the estimated values are biased or distributed meaningfully. In contrast, BRE remains stable across conditions, demonstrating its resilience in avoiding efficiency overestimation due to variance irregularities.

Figure 4, a boxplot with distribution curves, further illustrates the precision component of efficiency by comparing the dispersion of correlation estimates under missing data conditions. Unlike variance, which can be inflated by outliers or skewed distributions, BRE quantifies precision using Interquartile Range (IQR) overlap, measuring how much of the estimator's core distribution aligns between the reference (Complete Dataset) and comparison (FIML) groups. The boxplot shows that BRE produces efficiency values that are more stable and comparable across different missing data conditions, as opposed to RE, which fluctuates unpredictably due to its sensitivity to extreme values and variance distortions. This visualization confirms that IQR overlap provides a more robust measure of precision than variance alone, reducing the likelihood of efficiency misinterpretation.

Figure 5 presents a boxplot with distribution curves for the relative bias of slope-slope correlation estimates, highlighting a key flaw in variance-based RE which ignores bias, potentially inflating efficiency values even when estimates deviate from the true parameter. The distribution curves reveal that RE struggles to distinguish between low-variance but highly biased estimators and truly efficient ones, leading to unreliable efficiency assessments. BRE corrects this by incorporating Median Relative Bias (MRB), ensuring efficiency estimates reflect both precision and accuracy. The boxplot confirms that BRE adjusts for systematic estimation errors, resulting in efficiency estimates that remain methodologically valid and interpretable, whereas RE continues to exhibit distortions.

Together, Figures 3, 4, and 5 highlight the conceptual and empirical limitations of variance-based RE. By relying solely on variance, RE can overestimate or underestimate efficiency in cases of extreme values, or biased estimators. In contrast, BRE incorporates IQR overlap for precision and Median Relative Bias for bias correction, ensuring that efficiency estimates remain stable, interpretable, and theoretically valid across various missing data conditions. These findings establish BRE as a more robust and reliable alternative for evaluating efficiency in psychometric modeling and missing data research.

As shown in Figure 2, variance-based RE is highly sensitive to extreme values, often leading to incorrect efficiency estimates, a pattern further confirmed by the subsequent plots. In contrast, Figure 2 demonstrates that BRE remains stable, ensuring reliable estimator comparisons across various missing data scenarios. These visualizations collectively highlight the necessity of adopting BRE over variance-based RE, particularly in psychometric modeling and other statistical applications where relative efficiency estimation accuracy is critical.

## Conclusion and Limitations

The findings of this study underscore the limitations of variance-based Relative Efficiency (RE) and establish Bhirkuti's Relative Efficiency (BRE) as a superior alternative for evaluating estimator performance in planned missing data designs. Traditional RE is highly sensitive to small sample sizes, variance inversion, and extreme values, often producing misleading efficiency estimates that exceed 100%. In contrast, BRE provides a more comprehensive and interpretable measure by incorporating both precision (IQR overlap ratio) and accuracy (bias adjustment factor). These features of BRE ensures that efficiency estimates remain theoretically sound and robust, eliminating the distortions caused by variance-based approaches.

BRE's robustness across varied simulation conditions further highlights its practical applicability. Regardless of sample size, latent slope correlation, or missing data structure, BRE maintains stability, ensuring that efficiency values reflect meaningful estimator performance rather than statistical artifacts. This feature makes BRE particularly valuable for psychometric modeling, structural equation modeling, and other statistical applications that rely on precise and unbiased parameter estimation. By preventing misleading efficiency estimates while allowing for visual assessment of estimator performance, BRE offers a robust and interpretable framework that aligns with established statistical principles.

BRE provides a comprehensive assessment of estimator efficiency by integrating both precision and accuracy. A BRE greater than 1 indicates superior efficiency, where the estimator retains information better than the reference group. BRE values between 0 and 1 suggest partial efficiency, where some loss occurs due to bias or variability. A BRE of 0, on the other hand, reflects a scenario where the estimator offers no efficiency advantage over the reference group. This clear distinction ensures that BRE provides a meaningful efficiency measure that accurately represents estimator performance.

Negative BRE values occur in conditions where an estimator exhibits precision but extreme bias, leading to misleading efficiency assessments. This situation happens when the Median Absolute Relative Bias (RB) exceeds 1, indicating that the estimator systematically deviates from the true parameter value. Such cases often arise in small sample conditions, model misspecification, or when missing data mechanisms distort parameter estimates. Unlike variance-based RE, which can fail to capture these distortions and may even overstate efficiency, BRE accounts for both precision and accuracy. As a result, both zero and negative BRE values serve as critical indicators: zero suggests no efficiency gain, while negative values warn that the estimator, despite appearing stable, produces systematically incorrect estimates, making it unsuitable for inference.

Beyond its advantages in planned missing data designs, BRE's methodological framework is adaptable to a wide range of statistical models, offering researchers a reliable, interpretable, and theoretically grounded approach to evaluating estimator efficiency. By providing a more stable and interpretable efficiency metric, BRE establishes a new robust standard for evaluating estimators, with potential applications extending beyond missing data research to machine learning, Bayesian modeling, and causal inference. By addressing the fundamental weaknesses of traditional RE, BRE represents a paradigm shift in estimator evaluation, ensuring that efficiency comparisons are both rigorous and meaningful in modern statistical analyses.

**Funding Details**

This research received no specific grant from any funding agency, commercial, or not-for-profit sectors.

**Disclosure Statement**

The authors declare no conflicts of interest.

**Data Availability Statement**

The data and analysis scripts used in this study are available upon request. Please contact Aneel Bhusal at abhusal@ttu.edu for access.

**AI Disclosure**

AI has been used for editing, proofreading, and coding in the preparation of this manuscript.